\documentclass[twocolumn]{aastex61}

\usepackage{graphicx}
\usepackage{subfigure}
\usepackage{amsmath}
\usepackage{epsfig}

\newcommand{\refsec}[1]{Section~\ref{#1}}
\newcommand{\reffig}[1]{Figure~\ref{#1}}

\newcommand{\TNG}{\texttt{IllustrisTNG}}
\newcommand{\autoGMM}{\texttt{auto-GMM}}
\shorttitle{Kinematic Decomposition}
\shortauthors{Du et al.}

\begin{document}
\title{Kinematic decomposition of \texttt{IllustrisTNG} disk galaxies: morphology and relation with morphological structures}

\correspondingauthor{Min Du}
\email{dumin@pku.edu.cn}

\author{Min Du}
\affil{Kavli Institute for Astronomy and Astrophysics, Peking University, Beijing 100871, China}

\author{Luis C. Ho}
\affiliation{Kavli Institute for Astronomy and Astrophysics, Peking University, Beijing 100871, China}
\affiliation{Department of Astronomy, School of Physics, Peking University, Beijing 100871, China}

\author{Victor P. Debattista}
\affiliation{Jeremiah Horrocks Institute, University of Central Lancashire, Preston PR1 2HE, UK}

\author{Annalisa Pillepich}
\affiliation{Max-Planck-Institut f$\ddot{u}$r Astronomie, K$\ddot{o}$nigstuhl 17, D-69117 Heidelberg, Germany}

\author{Dylan Nelson}
\affiliation{Max-Planck-Institut f$\ddot{u}$r Astrophysik, Karl-Schwarzschild-Str. 1, 85741 Garching, Germany}

\author{Dongyao Zhao}
\affil{Kavli Institute for Astronomy and Astrophysics, Peking University, Beijing 100871, China}

\author{Lars Hernquist}
\affiliation{Harvard--Smithsonian Center for Astrophysics, 60 Garden Street, Cambridge, MA 02138, USA}

\begin{abstract}
We recently developed an automated method, \autoGMM\, to decompose simulated galaxies. 
It extracts kinematic structures in an 
accurate, efficient, and unsupervised way. We use \autoGMM\ to study the 
stellar kinematic structures of disk galaxies from the TNG100 run of \TNG. We 
identify four to five structures that are commonly 
present among the diverse galaxy population. Structures having 
strong to moderate rotation are defined as cold and warm disks, respectively.
Spheroidal structures dominated by random motions are classified as bulges or 
stellar halos, depending on how tightly bound they are. Disky bulges are 
structures that have moderate rotation but compact morphology. 
Across all disky galaxies 
and accounting for the stellar mass within 3 half-mass radii, the kinematic 
spheroidal structures, obtained by summing up stars 
of bulges and halos, contribute $\sim 45\%$ of the total stellar mass, while 
the disky structures constitute $\sim 55\%$. This study also provides important insights about the relationship 
between kinematically and morphologically derived 
galactic structures. Comparing the morphology of kinematic structures with that of 
traditional bulge+disk decomposition, we 
conclude: (1) the morphologically decomposed bulges are composite structures 
comprised of a slowly rotating bulge, an inner halo, and a disky bulge; (2) 
kinematically disky bulges, akin to what are commonly called pseudo bulges in 
observations, are compact disk-like components that have rotation similar to 
warm disks; (3) halos contribute almost $30\%$ of the surface density of the 
outer part of morphological disks when viewed face-on; and (4) both cold and 
warm disks are often truncated in central regions. 
\end{abstract}


\section{Introduction}

Classification of galaxies is one of the most fundamental and important steps in understanding galaxy evolution. The decomposition of galaxies into a bulge and a disk component serves as a foundation for the classification of galaxies into the Hubble (1926) sequence \citep{Sandage1981}, even if galaxies often comprise additional structures. For example, the Milky Way is a prototypical spiral galaxy exhibiting several stellar components, including a thin and thick disk, a boxy/peanut-shaped bulge, a bar, a stellar halo, and a nuclear star cluster \citep[see the review by][]{Bland-Hawthorn&Gerhard2016}. Many nearby galaxies that can be well-resolved in detail also have a thick disk that is old and metal-poor with respect to the thin disk \citep{Dalcanton&Bernstein2002, Yoachim2006, Comeron2011, Comeron2014, Elmegreen2017}. Meanwhile, the morphology of bulges covers a broad range, from highly spherically symmetric to flat \citep{Andredakis&Sanders1994, Andredakis1995, Courteau1996, Mendez-Abreu2010}. Classical bulges, likely the end-products of galaxy mergers \citep{Toomre1977}, are expected to be dynamically hot, spheroidal, and centrally concentrated. More flattened, rotationally supported spheroids, named pseudo bulges, are expected to be an outcome of internal secular processes \citep[e.g.,][]{Kormendy&Kennicutt2004}. The rich diversity of structures observed among nearby galaxies is evidence of the complex formation and evolutionary history of galaxies. To understand the formation and evolution of galaxies, accurate recognition and decomposition of structures is essential; this however is a very challenging task due to the incomplete information that can be inferred from observations and, in particular, to the confusion from line-of-sight projection. 

Galaxies can be better characterized and decomposed when information about their internal kinematics is observationally available. The rapid development of integral-field unit (IFU) spectroscopy \citep[e.g.,][]{Emsellem2007, Emsellem2011, Cappellari2011a, Cappellari2011b}, has led to rapid progress in decomposing galaxies with the aid of kinematical information. \citet{Zhu2018b} first applied the orbit-superposition method \citep[e.g.,][]{Schwarzschild1979, Valluri2004, vandenBosch2008} to reconstruct stellar orbits based on the stellar kinematics of galaxies in the CALIFA survey \citep{Sanchez2012}, and was therefore able to successfully extract kinematically cold, warm, hot, and counter-rotating components \citep{Zhu2018a, Zhu2018c}. Recently, \citet{Zhu2020} further included the stellar population distribution in modelling galaxies.

The development of kinematic decomposition helps to break the degeneracy in morphology of different structures identified by the distribution of stellar mass or light. For instance, contrary to expectations from morphological decomposition \citep[e.g.,][but see \citet{Gao2020}]{Fisher&Drory2008}, recent kinematic studies found no clear correlation between the S\'ersic index $n$ and galactic kinematic properties derived with the IFU technique \citep[e.g.][]{Krajnovic2013, Schulze2018}. \citet{Zhu2018c} also suggested that $n$ is not a good discriminator between rotating pseudo bulges and classical bulges with weak rotation. Therefore, it is crucial to elucidate the intrinsic relationship between kinematic structures and the more familiar, traditional morphological structures. 


Notwithstanding the advances enabled by IFU spectroscopy, the level of detail in which galaxies can be decomposed based on observations remains quite restrictive. Complementary progress can be made by turning to large-scale hydrodynamical cosmological simulations, which self-consistently capture the properties of the stellar and gaseous components of galaxies. As well-informed spectators, we can extract intrinsic structures in simulated galaxies, and track their evolutionary history. In recent years, significant progress has been made in modelling star formation and stellar feedback in simulations \citep{Agertz2011, Guedes2011, Aumer2013, Stinson2013b, Marinacci2014, Roskar2014, Murante2015, Colin2016, Grand2017}, to the point that galaxies with realistic bulge+disk structures can be reproduced in a fully cosmological context, including Illustris \citep{Vogelsberger2014a, Vogelsberger2014b, Genel2014,Nelson2015} and its follow-up project \TNG\ \citep{Marinacci2018, Naiman2018, Nelson2018a, Nelson2019a, Pillepich2018b, Pillepich2019, Springel2018}, EAGLE \citep{Schaye2015, Crain2015}, and Horizon-AGN \citep{Dubois2016}; see the review by \citet{Vogelsberger2019}. Galaxy zoom-in simulations, such as Auriga \citep{Grand2017}, FIRE \citep{Hopkins2014, Hopkins2018}, and NIHAO \citep{Wang2015}, have been able to generate galaxies with multiple structures beyond the simple bulge+disk components \citep[e.g.,][]{Brook2012b, Algorry2017, Ma2017, Obreja2018b, Gargiulo2019}. Recently, the highest resolution run of the \TNG\ (TNG50) simulated a fully representative cosmological volume of $(50\ {\rm Mpc})^3$ at a resolution comparable to that of zoom-in simulations \citep{Nelson2019, Pillepich2019}. In light of this development, a modern method that can decompose intrinsic structures in a physical way beyond the most basic bulge+disk components is required to make full use of the power of hydrodynamical cosmological simulations. 

\citet{Du2019} presented a fully automated method, \autoGMM, to identify different kinematic components of a galaxy accurately and efficiently. Gaussian mixture models in \autoGMM\ serve as an unsupervised machine learning algorithm to isolate distinct structures in the kinematic phase space of simulated galaxies \citep[see also][]{Obreja2016, Obreja2018a}. We automated the code to determine the number of Gaussian components allowed by the data through a modified Bayesian information criterion. As a result, the possibility of human bias is minimized using this method. As shown in \citet{Du2019}, auto-GMM successfully identifies kinematic structures in galaxies with diverse morphological and kinematic properties. 

Thanks to an updated galaxy physics model \citep{Weinberger2017, Pillepich2018a}, galaxies in the \TNG\ successfully capture many of the observed optical morphologies of nearby galaxies \citep{Huertas-Company2019, Rodriguez-Gomez2019}. The realism of the mock galaxies inspires confidence that the latest simulations can be used for detailed statistical studies and to inform the correlation between kinematically and morphologically defined galaxy components. In this work, we apply \autoGMM\ to a large sample of galaxies from the TNG100 run of the \TNG\ simulations. 

This paper is organized as follows. Our methodology is demonstrated in \refsec{methodology}. \refsec{sample} introduces the sample of disk galaxies used in our analysis. The two methods
used to classify structures found in kinematic phase space are described in \refsec{phasespace}. The basic morphological properties of such kinematic structures\footnote{The mass fraction and images of each kinematic structure are publicly available at \url{www.tng-project.org/data/}.} are shown in \refsec{result} and compared with the bulge+disk structures decomposed based on morphology. Our conclusions are summarized in \refsec{conclusion}.

\section{Methodology of kinematic decomposition}
\label{methodology}

\begin{figure*}[htbp]
\begin{center}
\includegraphics[width=0.75\textwidth]{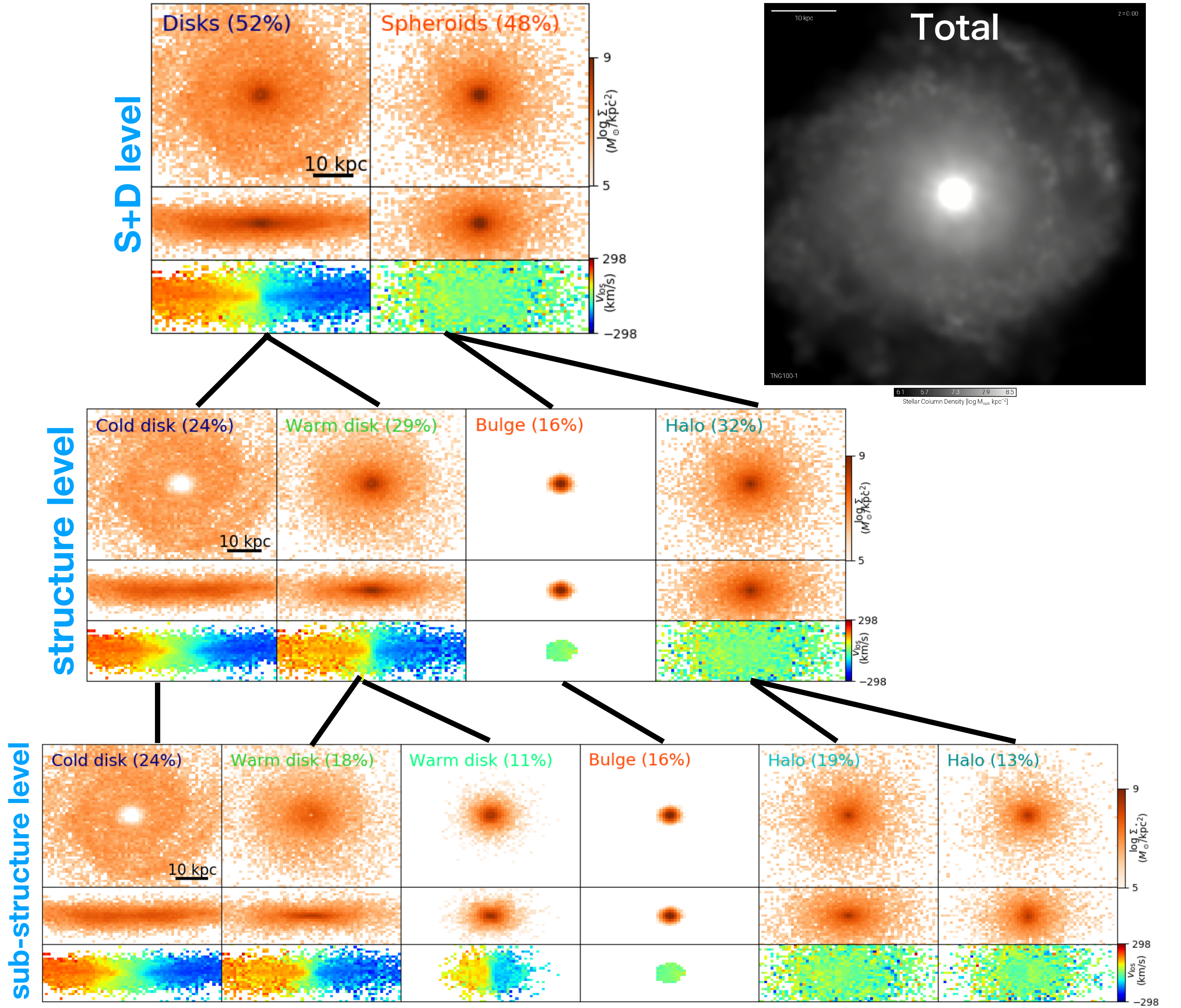}
\caption{An illustration of the hierarchical framework to understand galaxies with \autoGMM\ for a typical disk galaxy (ID 96505) in TNG100. The top-right panel shows the face-on smoothed stellar column density of all stars produced by the \TNG\ online visualization tool. The other three rows correspond to three levels of detail decomposed by \autoGMM: the spheroid+disk (S+D) level, the structure level, and the sub-structure level. Each row shows, from top to bottom, the face-on and edge-on views of the surface density distribution, and the edge-on mean velocity distribution. The structures at each lower level merge into the same structure in the higher level based on their similarities in kinematics. The structure classification used here uses classification method 1 described in \refsec{class1}. We regard the S+D level of the kinematic decomposition as the counterpart of the morphological bulge+disk decomposition. In this paper, we will not make use of the sub-structure level detail. Images of all disk galaxies are released on the \TNG\ website \url{https://www.tng-project.org/data/docs/specifications/}.}
\label{fig:levels}
\end{center}
\end{figure*}

\begin{figure*}[htbp]
\begin{center}
\includegraphics[width=0.95\textwidth]{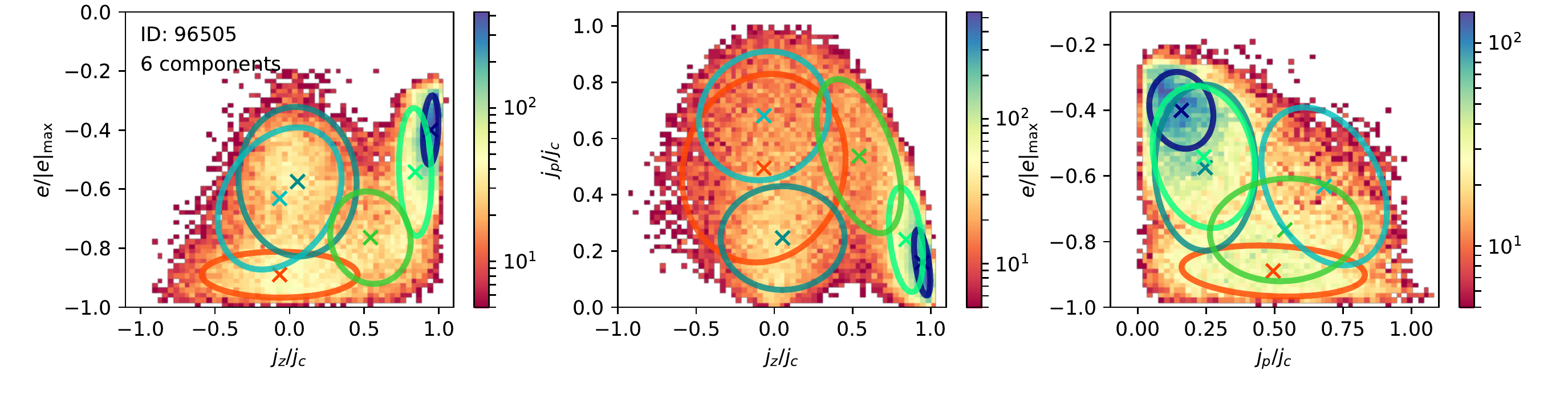}
\caption{The kinematic phase space ($j_z/j_c$, $j_p/j_c$, and $e/|e|_{\rm max}$) of the stars in the galaxy shown in \reffig{fig:levels}. The ellipses ($\sim 63\%$ confidence) correspond to the six Gaussian components (the sub-structure level in \reffig{fig:levels}) in the same color scheme. The crosses mark their centers. The colorbars represent the number of stellar particles in each bin.}
\label{fig:levels_jzjce}
\end{center}
\end{figure*}

It is natural that stars belonging to the same physical structure should cluster in their kinematic phase space. We characterize each star by three dimensionless parameters in kinematic phase space: the circularity parameter $\epsilon=j_z/j_c(e)$ \citep{Abadi2003b}, non-azimuthal angular momentum $j_p/j_c(e)$, and binding energy normalized by the minimum value $e/|e|_{\rm max}$, as proposed by \citet{Domenech-Moral2012}. The specific azimuthal angular momentum $j_z$ and non-azimuthal angular momentum $j_p$ are normalized by $j_c$, the maximum angular momentum having the same specific binding energy $e$. The maximum value of $|e|$ across the whole galaxy, $|e|_{\rm max}$, corresponds to the energy of the star that is most tightly bound at the galactic center. Thus, $j_z/j_c$ and $j_p/j_c$ quantify the aligned and misaligned rotation with the overall angular momentum, respectively, and $e/|e|_{\rm max}$ describes how tightly bound a stellar particle is. Consequently, the same standard is used to describe the kinematic properties of stars in diverse galaxies across the full mass range. All galaxies are oriented with their total angular momentum along the $z$-axis. We apply the code from \citet{Obreja2018a} to build the kinematic phase space of $j_z/j_c$, $j_p/j_c$, and $e/|e|_{\rm max}$ for all stars gravitationally bound to the galaxy with no limitation in galactocentric distance in individual galaxies, assuming every galaxy is isolated. 

We employ \autoGMM\ to identify structures through the clustering of the kinematic phase space parameters for all stars in each galaxy. \autoGMM\ is developed by combining the {\tt GaussianMixture} module from the \texttt{PYTHON} {\tt scikit-learn} package \citep{scikit-learn} with a modified Bayesian information criterion \citep[][see Section 2.3]{Du2019}. We briefly summarize this method here. The procedure is the same as that in \citet{Du2019}. The {\tt GaussianMixture} module is an unsupervised machine learning algorithm that clusters data efficiently to a target number of Gaussian components. Unlike most nonparametric clustering techniques that give ``hard'' assignments of stars to components, the parametric GMM approach allows ``soft''
probabilistic assignments. Each Gaussian component is a triaxial ellipsoid in the three-dimensional kinematic phase space. As recommended by \citet{Du2019}, we allow the number of Gaussian components to be determined automatically by the modified Bayesian information criterion ($\Delta$BIC), with the criterion $\Delta{\rm BIC}<C_{\rm BIC}$. We defined $\Delta$BIC as BIC-BIC$_{\rm min}$, where BIC$_{\rm min}$ corresponds to the ``ideal'' model having numerous Gaussian components. For the large data sample of stars in any galaxy, $\Delta$BIC is approximately equal to $-2{\rm ln}BF$. Here $BF$ is the Bayes factor \citep{Kass&Raftery1995} that can be used to quantify the evidence for the multiple Gaussian model auto-GMM found, compared to the perfect model. For two identical models, BF is equal to 1. Here we set $C_{\rm BIC} = 0.1$, which corresponds to $0.95 < BF < 1$; this is considered as an equally well fitting with respect to the ideal model. As shown in \citet{Du2019}, $C_{\rm BIC}$ is the only parameter that introduces a minor artificial effect. A reasonable $C_{\rm BIC}$ can vary in 0.05-0.15, resulting in slightly more or fewer Gaussian components. This criterion not only successfully avoids overfitting due to the use of too many components, but also minimizes the possibility of human bias. Its automated character enables implementation to a large sample of galaxies from cosmological simulations. 

Furthermore, \autoGMM\ serves as a framework for understanding galaxy structures. Realistic galactic structures inevitably contain some degree of finer sub-structure, not the least of which because the distribution function of structures in galaxies may not follow a simple Gaussian. Multiple Gaussian components, corresponding to sub-structures, are allowed to exist in the same kinematic structure. Sub-structures with similar kinematics contribute hierarchically to a higher level of structure. \reffig{fig:levels} illustrates an example galaxy (ID 96505) from the TNG100 run of \TNG\ at $z=0$. \autoGMM\ finds six kinematic Gaussian components at the sub-structure level (bottom panels). The positions of these Gaussian components on the kinematic phase space are shown in \reffig{fig:levels_jzjce}. The two disky components with a similar circularity parameter are considered as sub-structures of the same warm disk structure; likewise, the two ``halos'' at the sub-structure level belong to the same halo at the structure level. A less prominent Gaussian component induced by using a smaller $C_{\rm BIC}$ will merge with a more certain structure, reducing this artificial effect. The uncertainty in our statistic results due to $C_{\rm BIC}$ is thus negligible. Summing up all disky (cold and warm disks) and spheroidal structures recovers the traditional spheroid+disk decomposition (S+D level). 

This work focuses on the morphological properties of the principal structures in the stellar 
component of simulated galaxies; sub-structures included in each structure will not be considered 
further in the following analysis. The criteria used to classify structures will 
be described in detail in \refsec{phasespace}.

\begin{figure*}[htbp]
\begin{center}
\includegraphics[width=0.95\textwidth]{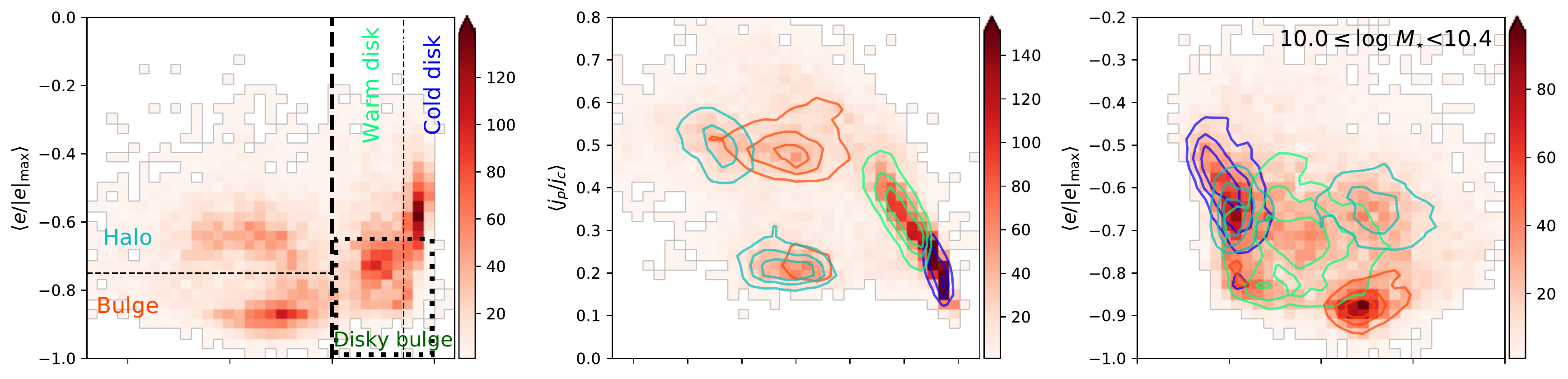}
\includegraphics[width=0.95\textwidth]{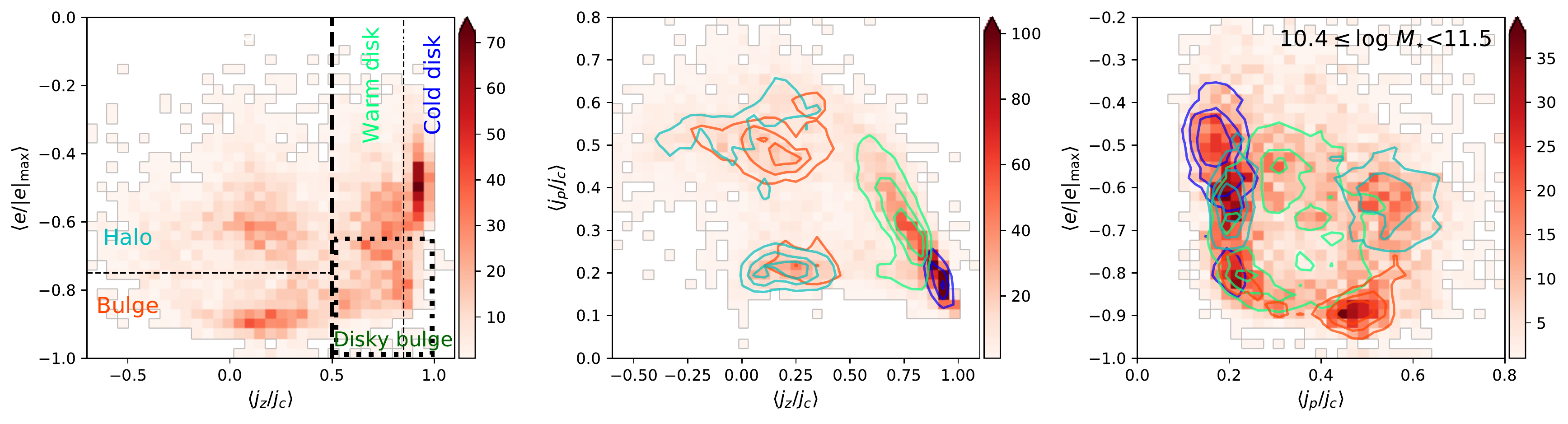}
\caption{The distribution of structural kinematic moments (average kinematics $\langle j_z/j_c \rangle$, $\langle j_p/j_c \rangle$, and $\langle e/|e|_{\rm max} \rangle$) of unbarred disk galaxies. Note that the kinematic moments here are the mass-weighted mean values of the three kinematical phase space parameters of all stars in each Gaussian component. Then the map represents the distribution of all Gaussian components decomposed by \autoGMM\ for the selected sample of galaxies instead of stars in a single galaxy. The galaxies are separated into mass ranges of (top) $10^{10.0}-10^{10.4}\,M_\odot$ and (bottom) $10^{10.4}-10^{11.5}\,M_\odot$. All components found by \autoGMM\ are classified into four kinds of structures in classification 1: cold disk (blue), warm disk (green), bulge (red), or halo (cyan). The contours in the same color show their distributions in maps of (middle) $\langle j_z/j_c \rangle$ vs. $\langle j_p/j_c \rangle$ and (right) $\langle j_p/j_c \rangle$ vs. $\langle e/|e|_{\rm max} \rangle$. The classification criteria are marked by the dashed lines. At the S+D level, they are classified into spheroids and disks by $\langle j_z/j_c \rangle=0.5$. At the structure level, spheroids are classified into halos and bulges by $\langle e/|e|_{\rm max} \rangle=-0.75$, while disks are classified into cold and warm disks by $\langle j_z/j_c \rangle=0.8$. The dotted square with $-1.0\leq\langle e/|e|_{\rm max} \rangle<-0.65$ and $0.5\leq\langle j_z/j_c \rangle\leq1.0$ marks the region of disky bulges defined in classification 2. }

\label{fig:kinemphase}
\end{center}
\end{figure*}

\begin{figure*}[htbp]
\begin{center}
\includegraphics[width=0.95\textwidth]{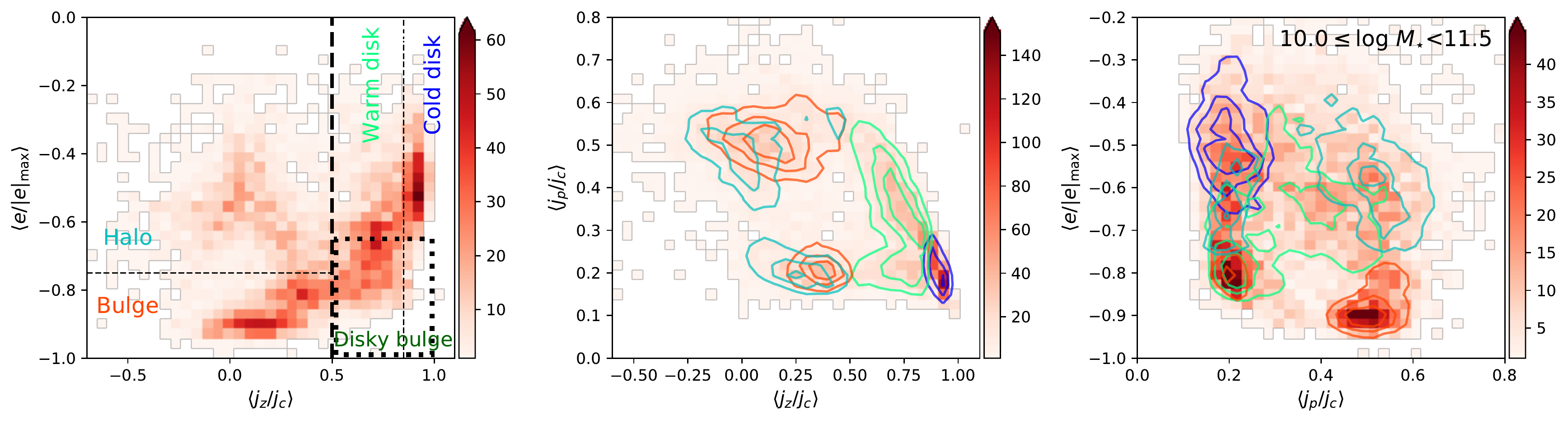}
\caption{The distribution of kinematic moments (average kinematics $\langle j_z/j_c \rangle$, $\langle j_p/j_c \rangle$, and $\langle e/|e|_{\rm max} \rangle$) of barred disk galaxies in the mass range $10^{10.0}-10^{11.5}\,M_\odot$. The classification criteria are the same as those used in \reffig{fig:kinemphase}.}
\label{fig:kinemphase_bar}
\end{center}
\end{figure*}

\section{Sample selection of disk galaxies}
\label{sample}

The \TNG\ Project \citep{Marinacci2018, Naiman2018, Nelson2018a, Nelson2019a, 
Pillepich2018b, Pillepich2019, Springel2018} is a suite of magneto-hydrodynamic
cosmological simulations run with the moving-mesh code {\tt AREPO} 
\citep{Springel2010, Pakmor2011, Pakmor2016}. \TNG\ has successfully 
reproduced many fundamental properties and scaling relations of observed 
galaxies. For example, the mass-size relation observed in both late-type and 
early-type galaxies has been well recovered within observational 
uncertainties \citep{Genel2018, Huertas-Company2019, Rodriguez-Gomez2019}. In 
particular, \citet{Rodriguez-Gomez2019} showed that the optical size and shape 
of the \TNG\ galaxies are consistent within $\sim 1\sigma$ scatter of the 
observed trends. Furthermore, \citet{XuDandan2019} found that the fractions of 
the different orbital components in \TNG\ are remarkably consistent with those 
estimated in CALIFA galaxies \citep{Zhu2018b}. Moreover, Illustris and \TNG\ 
have reproduced galaxies with unusual structural and kinematic properties, 
such as shell galaxies \citep{Pop2017,Pop2018}, low surface brightness
galaxies \citep{ZhuQirong2018}, and jellyfish
galaxies \citep{Yun2019}.  The great success of these simulations
gives us confidence that they can be used for statistical studies of kinematic 
structures and to obtain insights about the relationship between kinematically 
and morphologically identified structures. To obtain adequate statistics as well as 
a proper resolution, we 
analyze the TNG100 run. TNG100 follows the evolution of $2 \times 1820^3$ 
resolution elements within a periodic cube measuring $75h^{-1} \approx 110.7$ 
Mpc on a side, which translates to an average baryonic mass resolution element 
of $1.39 \times 10^6 \, M_\odot$. The gravitational softening length of the 
stellar particles is $0.5h^{-1} \approx 0.74$ kpc. 

To ensure that the galaxies have well-resolved structures, we concentrate on subhalos having stellar 
masses greater than $10^{10}\, M_\odot$, identified by using the Friends-of-Friends (FOF)
\citep{Davis1985} and SUBFIND \citep{Springel2001} algorithms. A total of 6507 
galaxies match this criterion. To compare galaxy properties to observations and
previous works, most of physical properties are measured within 3 times the 
three-dimensional half-mass radius, $r_e$. We verified that 
the statistic results measured for all stars or for stars within 30 kpc generally show
only minor difference from those measured within $3r_e$. We classify selected galaxies into 
disk and elliptical galaxies according to the relative importance of the 
kinetic energy in ordered rotation, $K_{\rm rot}=\langle v_\phi^2/v^2 \rangle$ 
\citep{Sales2010}, where $v_\phi$ and $v$ are the rotation velocity and total 
velocity, respectively, for each star. The quantity $K_{\rm rot}$ measures the 
mass-weighted average value of $v_\phi^2/v^2$  within a sphere of 30 kpc for 
each galaxy. In all, 3931 rotation-dominated galaxies with $K_{\rm rot}\geq 0.5$ are 
found; the other ones are classified as ellipticals. 

Stars moving on bar orbits generally have significant radial motions, which 
cause mixing in the kinematic phase space of $j_z/j_c$ and 
$j_p/j_c$ between disks and spheroids \citep{Du2019}. Thus, the presence of 
bars likely pollutes any analysis based on kinematic decomposition. Bar 
structures are commonly present in massive galaxies in TNG100. 
\citet{Rosas-Guevara2019} found bars in $37\%$ of disk galaxies with total 
stellar mass $M_{\star} \geq 10^{10.4}\, M_\odot$. In addition, D. 
Zhao et al. (2020, in preparation), applying the same criterion for selecting 
disk galaxies as we do, show that the bar fraction reaches $52\%$, consistent 
with the results of near-infrared surveys of nearby galaxies
\citep[e.g.,][]{Eskridge2000, Knapen2000, Marinova&Jogee2007, 
Menendez-Delmestre2007, Diaz-Garcia2016, Erwin2018}. A similar conclusion 
is reached independently by \citet{Zhou2020}. Based on the TNG100 
galaxy catalog of D. Zhao et al. (2020, in preparation), we further separate disk 
galaxies into sub-samples of barred and unbarred galaxies. Bar structures are 
identified with the observational criteria used in \citet{Marinova&Jogee2007}, based 
on ellipticity and position angle criteria imposed on isophotal analysis of face-on, 
mass-based surface density maps: (1) The maximum value of 
ellipticity is larger than 0.25, meanwhile the variation of ellipse position angle is 
less than 10 degrees; (2) ellipticity decreases noticeably outward from the maximum. 
Most galaxies with stellar mass $<10^{10.5} M_\odot$ host bars of half-size 
radius $<3$ kpc. More massive galaxies host longer bars, qualitatively consistent 
with the observed relation between bar size and galaxy stellar mass 
\citep{Erwin2018}, though TNG100 overproduces many relatively short bars. We use 
this subset to investigate the impact of bars on our main results.

\begin{figure*}[htbp]
\begin{center}
\includegraphics[width=0.98\textwidth]{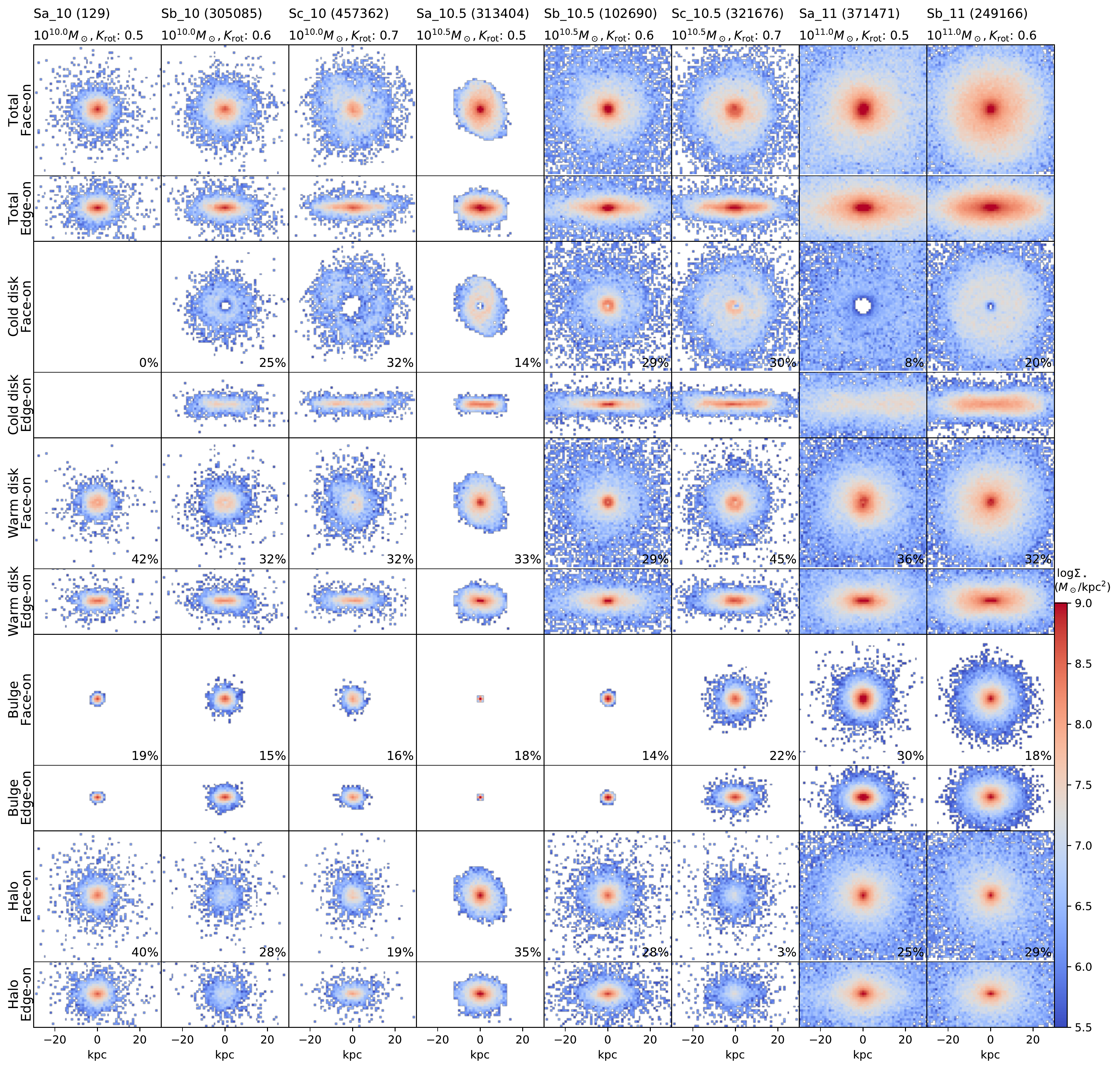}
\caption{Kinematic structures of example galaxies selected from TNG100 at $z=0$, classified according to method 1. From top to bottom, we show the face-on and edge-on surface density distributions of the entire galaxy, cold disk, warm disk, bulge, and halo. The bottom-right corner gives the mass fraction of each structure estimated from stars within $3r_e$. We show eight galaxies of different stellar mass ($M_\star = 10^{10}-10^{11}\,M_\odot$) and rotation ($K_{\rm rot}=0.5-0.7$); see the second row of the header for each column. Few galaxies of $M_{\star}=10^{11}\,M_{\odot}$ have strong rotation, and thus no example is found for $K_{\rm rot}=0.7$. The numbers in brackets are the ID of each galaxy in TNG100.} 
\label{fig:example2}
\end{center}
\end{figure*}

\section{Intrinsic Structures Found in The Distribution of Kinematic Moments}
\label{phasespace}
 
We construct the distribution of kinematic moments for the Gaussian components 
obtained from applying \autoGMM\ to all disk galaxies selected from TNG100. 
The structural kinematic moments are the mass-weighted mean values of 
the three kinematical phase space parameters of all stars gravitationally bound to the galaxy
in each Gaussian component, $\langle j_z/j_c \rangle$, $\langle j_p/j_c \rangle$, and 
$\langle e/|e|_{\rm max} \rangle$.

Intrinsic structures should naturally cluster in structural kinematic moments,
in a similar way that stars do in an individual galaxy. In this section, we 
introduce two methods to classify the \autoGMM\ components by their structural 
kinematic moments. 

\subsection{Classification 1: cold/warm disk, bulge, and halo}
\label{class1}

Figures \ref{fig:kinemphase} and \ref{fig:kinemphase_bar} show the structure 
kinematic phase spaces of unbarred and barred disk galaxies, respectively. 
Unbarred galaxies are separated into two mass bins, $M_{\star}=
10^{10.0}-10^{10.4}\,M_\odot$ and $10^{10.4}-10^{11.5}\, M_\odot$. Four distinguishable 
clusters, likely corresponding to intrinsic structures, are clearly seen.
These features are not significantly affected by the stellar mass of the 
galaxies or whether they host a bar. 

Every \autoGMM\ component can be easily classified into spheroidal or disky structures
by setting a circularity criterion $\langle j_z/j_c \rangle=0.5$ 
(thick dashed line). We further classify spheroidal components into bulges and 
halos by the criterion $\langle e/|e|_{\rm max} \rangle=-0.75$ (horizontal 
dashed line), while the disky components are classified into cold or warm 
disks by $\langle j_z/j_c \rangle=0.85$. These structures also cluster well by 
introducing the $\langle j_p/j_c \rangle$ dimension (see the contours in the 
right two panels). In this scheme, which we call classification 1, every 
galaxy is deconstructed at the structure level or the spheroids+disks (S+D) 
level (\reffig{fig:levels}). Classification 1 uses the same criteria as those 
in \citet{Du2019}. 
In previous studies \citep[e.g.][]{Abadi2003b}, the bulge is commonly defined as a structure without rotation, in which star particles distribute asymetrically around $j_z/j_c=0$. However, the physical justification is debatable. \reffig{fig:kinemphase} shows that bulge structures cluster around $\langle j_z/j_c \rangle=0.2$, 
indicating weak rotation. The assumption of no rotation may significantly underestimate spheroids.

A similar pattern emerges in the structure kinematic phase space of barred 
galaxies (\reffig{fig:kinemphase_bar}). However, there is no clear boundary 
that separates warm disks and bulges in barred galaxies, possibly due to the 
kinematic mixture of stars in the bar. Our adopted kinematic phase space 
cannot distinguish ordered radial motions, resulting in the misclassification 
of some dynamically cold stars into a hotter structure. 

\reffig{fig:example2} shows a few examples of galaxies with 
$M_{\star}= 10^{10}$, $10^{10.5}$, and $10^{11}\,M_\odot$ and 
$K_{\rm rot}=0.5, 0.6$, and 0.7, randomly selected from the parent sample. 
From left to right, they can be separated into three groups by mass. Selected 
galaxies with $K_{\rm rot}=0.5, 0.6,$ and $0.7$ are named Sa\_*, Sb\_*, and 
Sc\_*, respectively, where $*$ corresponds to the logarithm of its stellar 
mass. The visual morphologies of their structures are shown in face-on (upper 
panels) and edge-on (lower panels) views. The top two rows represent the 
surface density distribution of all stars. It is clear that, with the increase 
of $K_{\rm rot}$ (from left to right in each group), galaxies having the same 
mass become more disky and less compact. 

The kinematic structures identified by classification 1 are shown from the 
third row to the bottom in \reffig{fig:example2}: cold disks, warm disks, 
bulges, and halos. It is clear that, from a morphological perspective, they 
are qualitatively consistent with thin disks, thick disks, bulges, and halos. 
The bottom-right corner of each panel provides the mass ratio of the 
corresponding structure measured within $3r_e$. The cold disks are the 
thinnest disky structure within each galaxy, and commonly host spiral arms (e.g.,
Sa\_10.5 and Sc\_10.5). The warm disks are also flattened, but vertically thicker 
than thin disks. Another clear difference between them is that warm disks are 
likely to be centrally concentrated, while some cold disks have an obvious 
central break (e.g., Sb\_10, Sa\_10, and Sa\_11). Both bulges and halos are 
slowly rotating spheroids. Our models thus confirm that dynamically cold stars 
in galaxies with high $j_z/j_c$ always form a flattened, disky structure, and 
that dynamically hot stars with low $j_z/j_c$ form a spheroidal structure. 
Kinematic halos extend into the galaxy center. This implies that bulges 
decomposed from their morphology should be the superposition of intrinsic 
bulges and halos. 

\begin{figure*}[htbp]
\begin{center}
\includegraphics[width=0.98\textwidth]{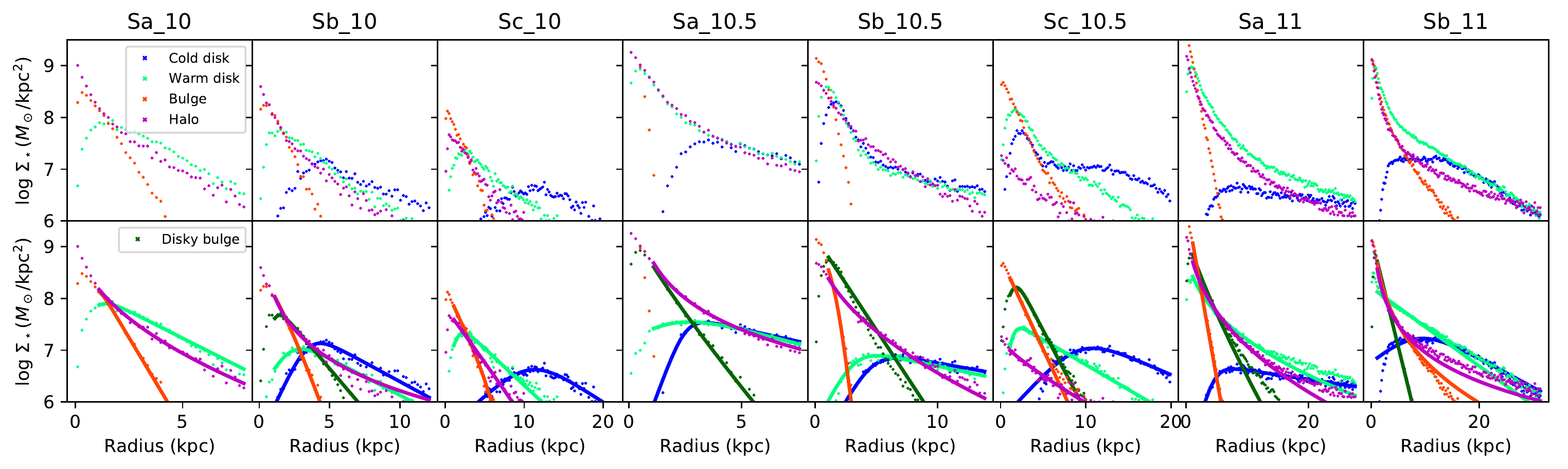}
\caption{Face-on surface density of the kinematic structures found by \autoGMM. The models correspond to the same galaxies shown in Figure \ref{fig:example2}. The top and bottom panels show the results obtained using classifications 1 and 2, respectively. The points are the surface density viewed face-on, averaged within annuli of width 0.2 kpc. The solid profiles in the bottom panels represent the model fit using either a centrally truncated exponential or a S\'ersic function. Note that the difference between classifications 1 and 2 arise from whether or not the disky bulge is considered as a separate structure from the cold and warm disks.}
\label{fig:example_1Dfaceon}
\end{center}
\end{figure*}

\begin{figure*}[htbp]
\begin{center}
\includegraphics[width=0.98\textwidth]{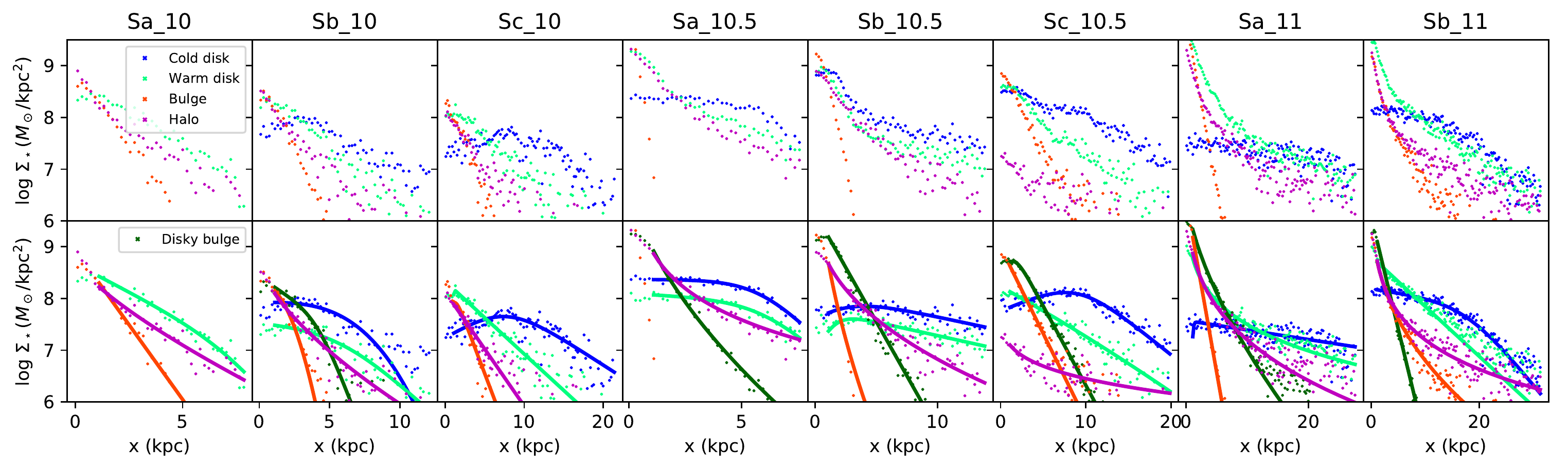}
\caption{Edge-on surface density profiles of the kinematic structures found by \autoGMM. The models correspond to the same galaxies shown in Figures \ref{fig:example2} and \ref{fig:example_1Dfaceon}. The top and bottom panels show the results obtained using classifications 1 and 2, respectively. The discrete data are measured from stars within rectangular bins of width $\Delta x = 0.2$ kpc and height $|z| \leq 0.5$ kpc along the major axis in edge-on view. The solid profiles in the bottom panels represent the best-fitting model using either a centrally truncated exponential or a S\'ersic function. }
\label{fig:example_1Dedgeon}
\end{center}
\end{figure*}

To better quantify the morphology, we measure the one-dimensional stellar surface 
density profiles for each structure obtained by classification 1. The top 
panels of Figures \ref{fig:example_1Dfaceon} and \ref{fig:example_1Dedgeon} 
show the face-on and edge-on views of the galaxy, respectively. As expected, 
both cold (blue) and warm disks (green) have somewhat exponential profiles.
Bulges (red) are compact, and halos (magenta) extend further out to large 
radius following a nearly S\'ersic function. However, the morphologies of 
disky structures, especially warm disks, are still quite complex using this 
classification. Cold disks are largely truncated in their central regions, 
while many warm disks seem to host an additional component that is much more 
compact than their outer regions (e.g., Sa\_10.5, Sb\_10.5, Sc\_10.5, Sa\_11, 
and Sb\_11). An extra compact component associated with the warm disk can 
also clearly be seen in the two-dimensional face-on images of these models 
(\reffig{fig:example2}). Thus, there is no simple way that the morphologies of 
the kinematic structures can be described with this classification method.


\begin{figure*}[htbp]
\begin{center}
\includegraphics[width=0.98\textwidth]{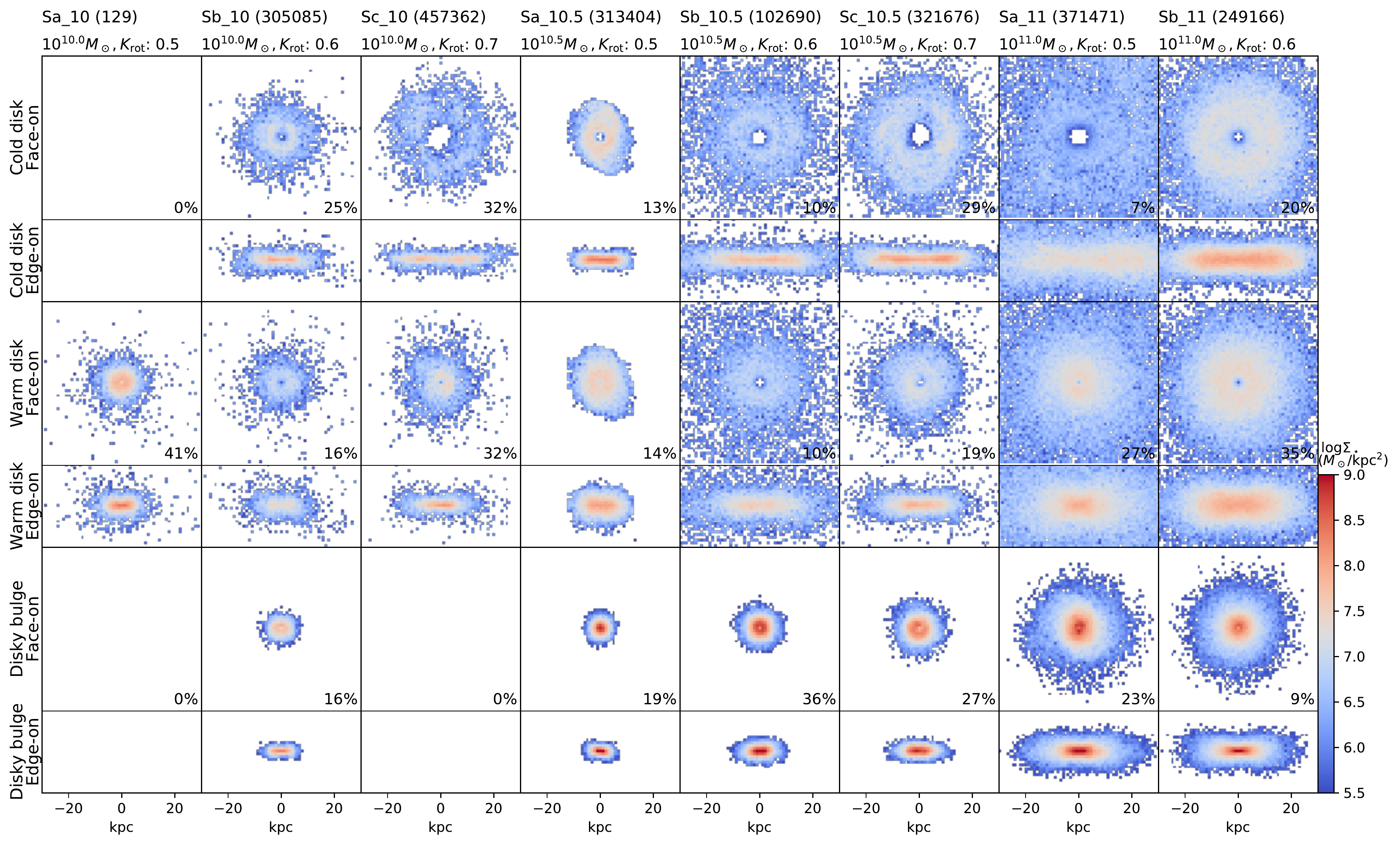}
\caption{Kinematic structures of examples selected from TNG100, classified according to method 2; the galaxies are the same as those shown in \reffig{fig:example2}. From top to bottom, we show the face-on and edge-on views of the surface density distributions of the cold disk, warm disk, and disky bulge. The mass ratios of each structure, labeled in the bottom-right corner of each panel, are estimated from stars within $3r_e$.}
\label{fig:example4}
\end{center}
\end{figure*}

\subsection{Classification 2: including a disky bulge}
\label{class2}

The central concentrations, common in many warm disks and in some cold disks, 
strongly suggest that our classification method 1 has failed to isolate an 
important, additional component. This additional component---characterized by
moderate rotation and a centrally concentrated morphology---motivates us to add
another structure that is likely located at the bottom-right corner of the 
$\langle j_z/j_c \rangle$ vs. $\langle e/|e|_{\rm max} \rangle$ diagram (marked
by the dotted square in Figures \ref{fig:kinemphase} and 
\ref{fig:kinemphase_bar}). 
We name this new structure the ``disky bulge'', and we designate this method of 
\autoGMM\ classification as classification 2. We show the one-dimensional 
surface density profiles of the structures identified by classification 2 in 
the bottom panels of Figures \ref{fig:example_1Dfaceon} and 
\ref{fig:example_1Dedgeon}, for comparison with those identified by 
classification 1. The bulge and halo structures are exactly the same in the 
two methods, while the extra central concentrations of cold and warm disks in 
classification 1 are assigned as disky bulges under classification 2. 
Removing the central concentration simplifies the density distributions of 
both the cold and warm disks. The disky bulge component accounts for roughly 
half of the mass of the warm disk. The two-dimensional stellar surface density 
distributions of these disky structures are shown in \reffig{fig:example4}. 
The flatten morphology and moderate rotation of disky bulges (bottom panels 
of \reffig{fig:example4}, see Section \ref{sec:kimdecom} for the statistic results) 
strongly suggest that they are the counterparts of the so-called pseudo bulges 
frequently observed \citep[e.g.,][]{Kormendy&Kennicutt2004}. 

It is worth emphasizing that the differences between warm disks and 
disky bulges are not always significant. In Figure \ref{fig:kinemphase}, there is no 
clear gap between the groups of the disky bulge and the warm disk. Although the 
separation of the disky bulge from disks helps to gain more freedom for comparing with 
morphological decompositions, it is unclear whether they form through a different mechanism.   
This topic lies beyond the scope of this paper.

\subsection{Fitting the kinematic structures of classification 2}
\label{sec:fitting}

As shown in the bottom panels of Figures \ref{fig:example_1Dfaceon} and 
\ref{fig:example_1Dedgeon}, all kinematic structures derived from 
classification 2 seem to follow regular morphologies that can be well-described
as a S\'ersic or truncated exponential function. The exponential 
function can be truncated by a hyperbolic function \citep{Peng2010}

\begin{equation}
T = 0.5 \left({\rm tanh} \left[(2-B)\frac{R}{R_b} + B\right] + 1\right),
\end{equation}

\noindent
where $B = 2.65 - 4.98R_b/(R_b - R_s)$, with $R_b$ the break radius and $R_s$ 
the softening in the cylindrical coordinate system. Only stars within 3 half-mass radii 
are used. The minimum density of each bin is limited to 
$10^6\,M_\odot\,{\rm kpc}^{-2}$ (about one stellar particle per 1 kpc$^2$). 
We also ignore the region $R\leq 1$ kpc in the fitting, in order to avoid the 
over-smoothing effect in the central region whose size is similar to the 
softening length (0.7 kpc) of the stars.

\begin{figure}[htbp]
\begin{center}
\includegraphics[width=0.45\textwidth]{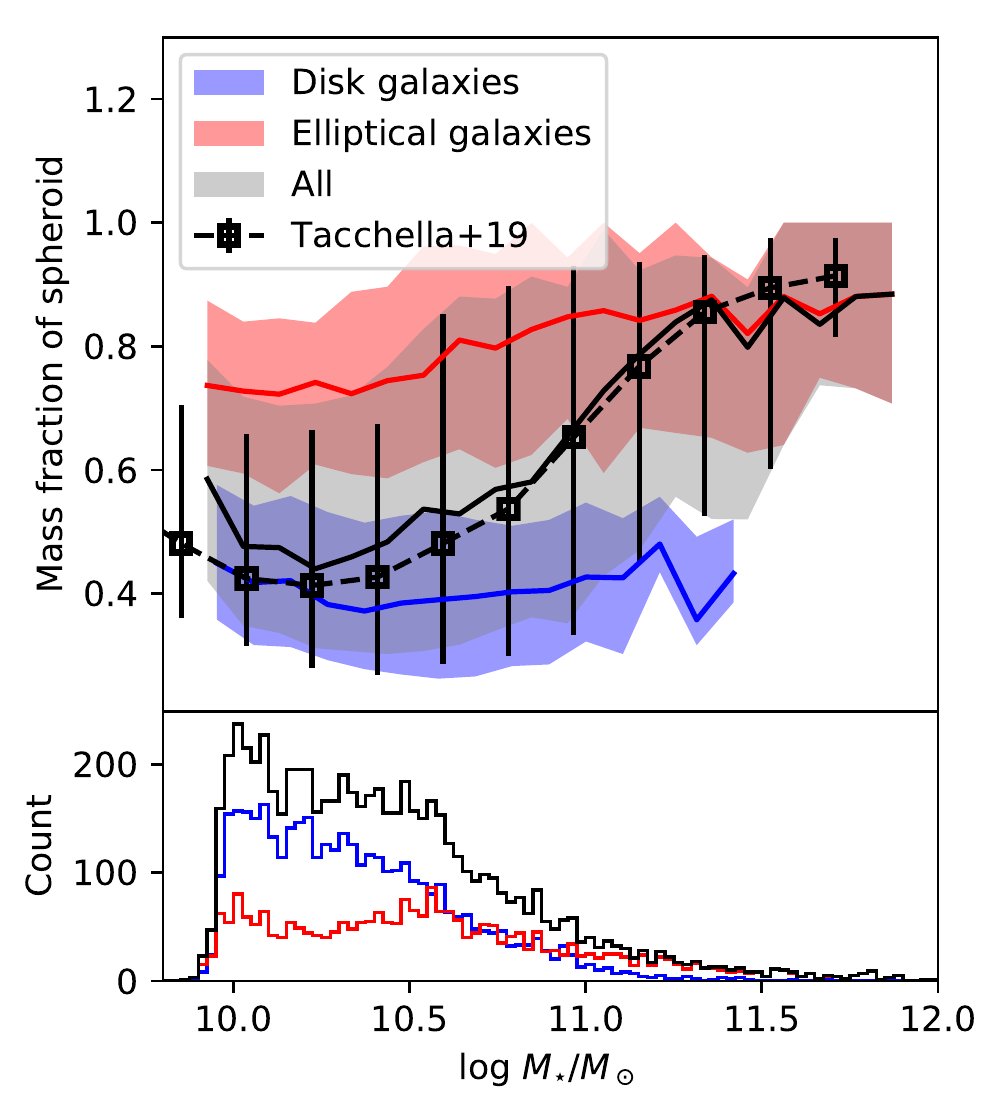}
\caption{Mass fractions of spheroidal structures in TNG100 galaxies. All galaxies of stellar mass $\geq 10^{10}\,M_\odot$ are included. The parent sample is classified into disk and elliptical galaxies by the criterion $K_{\rm rot}=0.5$. The solid profiles represent the median, while the shaded regions are the envelopes of $1\sigma$ scatter. The spheroids include the bulges and halos identified kinematically by \autoGMM. Here, $M_{\star}$ and mass fraction are measured from stars within $3r_e$, to facilitate comparison with previous results and observations. The mass fractions of spheroids obtained by \autoGMM\ are consistent with the results of \citet{Tacchella2019}. The number count of galaxies in each mass bin is shown in the bottom panel.}
\label{fig:mratio1}
\end{center}
\end{figure}

The fit results (solid profiles) of the selected examples are overlaid in the bottom panels of Figures \ref{fig:example_1Dfaceon} and \ref{fig:example_1Dedgeon}. The best-fitting function is selected automatically based on $\chi^2$, defined as 

\begin{equation}
\chi^2 = \frac{1}{N}\sum_{i=0}^N\left(\frac{\Sigma_{\star, i} - \Sigma_{{\rm model}, i}}{\Sigma_{\star, i}}\right)^2, 
\end{equation}

\noindent
where $\Sigma_{\star, i}$ and $\Sigma_{{\rm model}, i}$ are the average surface density of the data and model, respectively. It is clear that all structures can be well-fitted, suggesting that the kinematic structures, when converted into radial mass surface density profiles, can be robustly captured by the functional forms traditionally used for photometric morphological components.

\section{Statistical properties of kinematic structures}
\label{result}

Kinematic parameters that quantify the relative importance of circular rotation have been 
widely used to decompose spheroidal components from disks in simulated galaxies 
\citep[e.g.,][]{Sales2010, Tacchella2019}. However, the relation between morphological and 
kinematic measurements is still fairly unclear, possibly leading to contradictory conclusions. 
For example, \citet{Huertas-Company2019} found that the morphologies of both high-mass 
and low-mass galaxies in TNG100 are in tension with visual morphologies of observed nearby galaxies.
In particular, they found that an overabundance of late-type galaxies ($\sim50\%$ versus $\sim20\%$) 
at the high-mass end ($M_\star > 10^{11}\,M_\odot$).
In contrast, \citet{Tacchella2019} concluded that both the mass fraction and concentration of spheroidal 
components identified by kinematics agree fairly well with observations. 

In this section, we study the statistical properties of intrinsic kinematic 
structures identified by \autoGMM. To link kinematic and morphological 
structures, we also decompose the disk galaxies morphologically, adopting a 
simple model comprising an exponential function for the disk and a S\'ersic 
function for the bulge, as is commonly applied to observed galaxy images. 
We use the mass-based one-dimensional face-on surface density profiles. 
Thus, the effects of dust and the change of mass-to-light ratio are not considered, 
same as the kinematic decomposition. In observations, the effects of dust on 
galaxy morphology is generally minimized via selecting dust-free galaxy samples 
or using redder band passes \citep[e.g.][]{Salo2015, Gao2020}. Again, we ignore the central 
region ($R\leq1$ kpc) due to the low resolution. This model works adequately 
for most of our galaxies.

\subsection{Mass fraction}
\label{sec:mratio}

\begin{figure}[htbp]
\begin{center}
\includegraphics[width=0.46\textwidth]{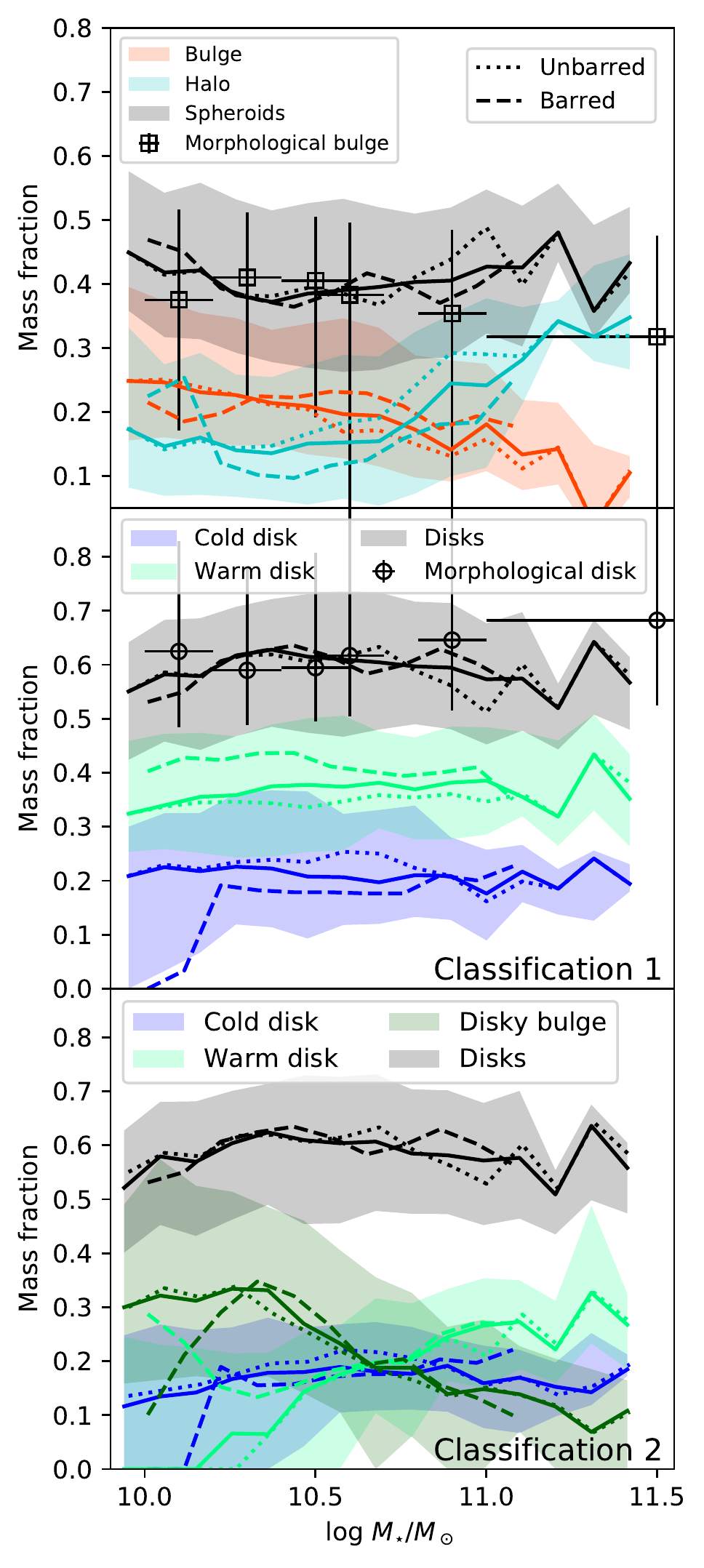}
\caption{Mass fractions of kinematic structures identified by \autoGMM\ in TNG100 disk galaxies. The solid profiles represent the median, while the shaded regions are the envelopes of $1\sigma$ scatter. Both $M_{\star}$ and mass fraction are measured from stars within $3r_e$. The results of disky structures obtained by classifications 1 and 2 are shown in the middle and bottom panels, respectively. The dotted and dashed profiles represent the medians of unbarred and barred galaxies, respectively. Barred galaxies are likely to host more bulges and warm disks, and thus less massive halos and cold disks. The squares and circles with error bars mark the mass fractions of bulges and disks, respectively, obtained from simple bulge+disk decomposition of the mass distribution.}
\label{fig:mratio2}
\end{center}
\end{figure}

Based on the hierarchical framework illustrated in \reffig{fig:levels}, we can 
easily estimate the properties of the kinematic structures. The mass fraction 
of the spheroid of each galaxy is given in the top panel of 
\reffig{fig:mratio1} by summing up the masses of the bulge and halo.
The red, blue, and black solid profiles represent, respectively, the medians 
of elliptical\footnote{We have also applied \autoGMM\ to decompose elliptical 
galaxies, even though they are generally featureless.}, disk, and all galaxies 
of $M_{\star}\geq10^{10}\, M_\odot$. The shaded regions indicate 
their $1\sigma$ scatter. The mass fraction of spheroids (black solid profile) 
clearly rises from 0.5 to 0.9 with increasing galaxy mass. This change is 
consistent with the sharp drop in the number of disk galaxies at 
$M_{\star}\approx 10^{10.6}\,M_\odot$, as shown in the bottom panel.
Both the trend and scatter of spheroid mass fraction are perfectly consistent 
with the results of \citet{Tacchella2019} (dashed line with error bars) who 
also adopted a kinematic approach. This trend 
is also consistent with the photometric bulge+disk decomposition of 7500 
local galaxies in the GAMA survey \citep{Moffett2016}, as suggested by 
\citet{Tacchella2019}. Similarly, \citet{Park2019} showed that 
kinematically-derived spheroids contribute to 43\% of total stellar mass 
in 144 field galaxies at $z=0.7$ in the New Horizon simulation 
(a zoom-in simulation of Horizon-AGN, Dubois et al. in preparation). Taking our sample of 
disk galaxies as a whole, spheroids contribute $\sim 40\%-50\%$ 
(the blue profile in the top panel of Figure \ref{fig:mratio1}),
roughly independent of stellar mass for the mass range considered here. 

\reffig{fig:mratio2} explores the mass fractions of the various kinematic 
structures in greater detail. The difference between barred and unbarred 
galaxies can be seen by comparing the dashed and dotted profiles. It is clear 
that the mass fraction of bulges is lower in less massive galaxies 
($M_{\star} \lesssim 10^{10.7}\,M_\odot$). The stellar halo mass 
fraction becomes more prominent in massive galaxies, possibly due to the 
increase of diffuse spheroids produced in dry mergers, as the amount of 
accreted stellar mass increases significantly in galaxies above 
$\sim10^{10.5}\,M_\odot$ \citep{Rodriguez-Gomez2016, Pillepich2018b}. Our two 
classification schemes produce exactly the same bulges and halos, their main 
differences manifesting only in terms of the cold and warm disks. 

The disky structures identified through our kinematic method contribute a 
constant mass fraction in the disk galaxies. Using classification 1, the cold 
disk comprises 20\%, with a scatter of $\pm 10\%$. 
The average mass fraction of warm disks is higher than that of cold disks by 
$75\%$. The mass fractions of both cold and warm disks are almost independent 
of galaxy stellar mass. The cold and warm disk mass fraction we obtain are 
similar to those extracted in CALIFA galaxies using the Schwarzschild method, 
for which \citet{Zhu2018a} estimate $\sim 20\%$ and $40\%$, respectively.
However, it is worth mentioning that \citet{Zhu2018a} measured luminosity 
fractions within $1r_e$. Over the same scales, \autoGMM\ obtains cold disk 
fractions $\lesssim 10\%$, pointing to possible systematic discrepancies 
between these two methods. Alternatively, stars are overheated in the central regions 
in the TNG100 galaxies, due to its low resolution. We intend to investigate 
this issue in the future using higher-resolution runs of \TNG\ (i.e. TNG50).

Classification 2 yields a roughly constant cold disk mass fraction of $15\%$, somewhat lower than that estimated by classification 1. The warm disk mass fraction rises with total galaxy mass, from nearly 0\% to about $30\%$.

The trends for bulges and halos are the same using classification 2. The frequency of both cold and warm disks decreases significantly in low-mass galaxies, as a consequence of the separation of disky bulges from cold and warm disks, leading to the significant decrease of warm disk mass fraction and the large scatter of cold disk mass fraction toward low-mass galaxies (bottom panel of \reffig{fig:mratio2}).

\begin{figure*}[htbp]
\begin{center}
\includegraphics[width=0.95\textwidth]{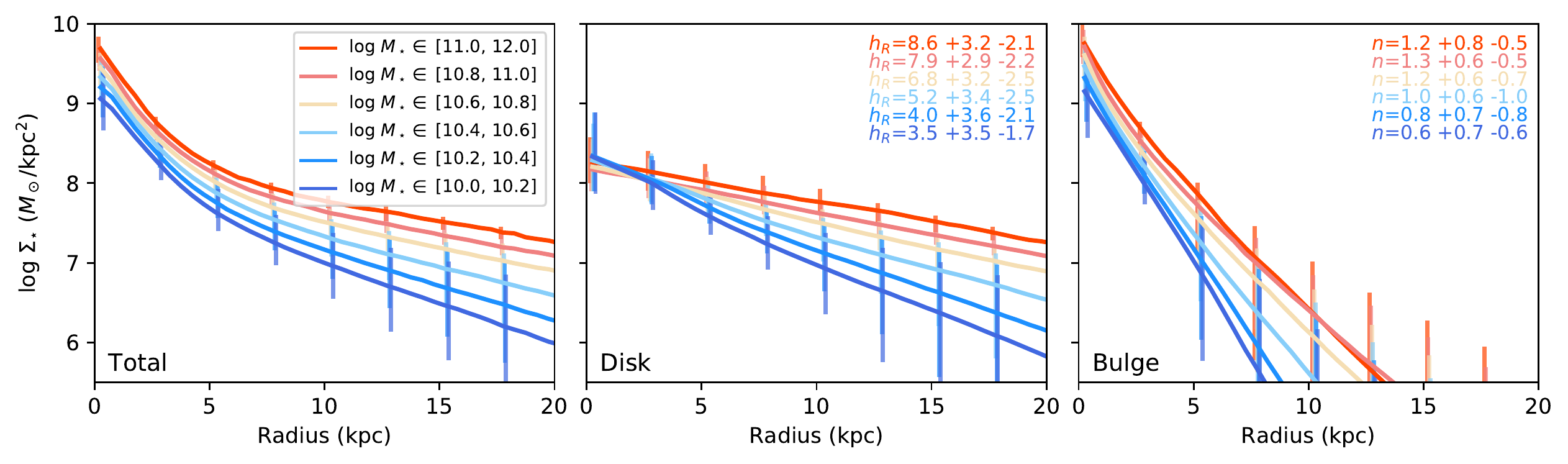}
\caption{Radial surface density profiles of $z=0$ disk galaxies from TNG100. We separate the galaxies into equal mass bins of 0.2 dex, except for the most massive ones. From left to right, the panels show the median surface density profiles of the entire galaxy, the exponential disk component alone, and the S\'ersic bulge component alone, decomposed by the morphological method. The error bars represent the $1\sigma$ scatter. To avoid overlap, we shift the error bars slightly for clarity.}
\label{fig:SurfaceDensity}
\end{center}
\end{figure*}

\begin{figure*}[htbp]
\begin{center}
\includegraphics[width=0.95\textwidth]{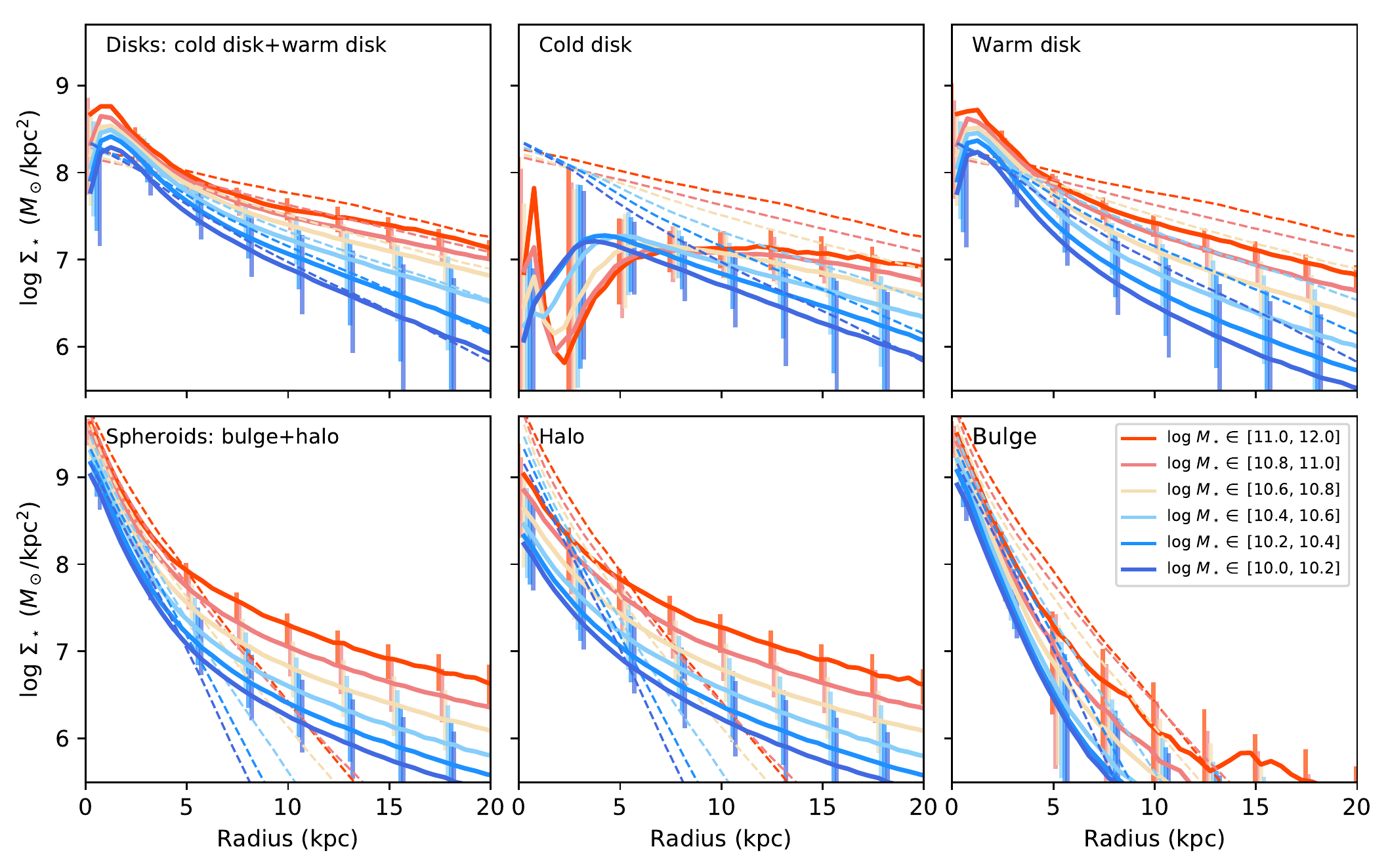}
\caption{Radial surface density profiles of kinematic structures decomposed by \autoGMM\, in $z=0$ disk galaxies from TNG100. Four structures are identified by classification 1. The spheroids are obtained by summing up bulges and halos, while the combination of cold and warm disks are the disks. The sample of disk galaxies is the same as that shown in \reffig{fig:SurfaceDensity}, separated into the equal mass bins. The error bars (slightly shifted for clarity) represent the $1\sigma$ scatter. The dashed median profiles of the morphological disks (top panels) and bulges (bottom panels) are overlaid to compare with the results of the kinematic decomposition.}
\label{fig:SurfaceDensity2}
\end{center}
\end{figure*}

The mass fractions of bulges and disks derived from traditional morphological
decomposition (squares and circles in \reffig{fig:mratio2}) are consistent with 
the mass fractions decomposed kinematically. However, the kinematically 
decomposed spheroids also include halos, whose morphologies are likely to be 
different from bulges. The kinematic bulges alone comprise only half or less 
of the mass fraction of morphological bulges. Morphologically identified
bulges cannot be linked directly to either kinematic spheroids or bulges. Thus, 
while the observed bulge+disk morphology is generally indeed an indicator of 
the underlying stellar kinematics, it is still unlikely to be an accurate proxy of the
diverse kinematic structures that are mixed in galaxies. In order to link the morphological 
and kinematic structures, their density profiles are compared in detail in the next section.
%

Compared to their unbarred counterparts, barred galaxies have $\sim 5\%-10\%$ 
higher mass fractions in warm disks and bulges. This is partially due to 
the misclassification of stars moving on bar orbits \citep{Du2019}, although
genuine differences between the two galaxy types cannot be excluded. For the 
purposes of this study, we simply note that the differences between barred 
and unbarred galaxies are smaller than their $1\sigma$ scatter and do not 
affect any of our main results.

\subsection{Radial surface density profiles}
\label{sec:SD}

To better understand the relation between the kinematic structures found by 
\autoGMM\ and morphological structures, we compare the radial density 
distributions of the kinematic structures with the bulges and disks derived 
from traditional morphological decomposition. 

\subsubsection{Morphological decomposition: bulge+disk}

\reffig{fig:SurfaceDensity} shows the one-dimensional surface density profiles 
as a function of radius. The panels, from left to right, give the face-on 
surface density profiles of the entire galaxy, the exponential disk alone, and 
the S\'ersic bulge alone. 

The disks have reasonable scale-lengths that increase with galaxy mass, from 
$h_R = 3.5^{+3.5}_{-1.7}$ kpc to $7.9^{+2.9}_{-2.2}$ kpc. This trend is 
roughly consistent with observations of galaxies over the stellar mass range 
$10^{10} -10^{11}\,M_\odot$ \citep[e.g.,][]{Fathi2010}. The S\'ersic index of 
the bulge component increases from $n = 0.9^{+0.7}_{-0.4}$ in lower mass,
 $M_{\star} = 10^{10}\, M_\odot$ galaxies to $n = 1.4^{+0.6}_{-0.4}$
in galaxies of $M_{\star} = 10^{11}\, M_\odot$. A S\'ersic index of 
$n < 2$ is generally used to separate pseudo bulges from classical ones 
\citep[e.g.,][but see \citet{Gao2020}]{Fisher&Drory2008}. Very few bulges with $n>2$ 
exist in TNG100 disk galaxies. Fitting the overall surface luminosity with a 
single S\'ersic function, \citet{Rodriguez-Gomez2019} have shown that TNG100 
generates lower values of $n$ with respect to Pan-STARRS observations of 
nearby galaxies. Our results reaffirm this trend and extend it explicitly 
to the scale of the bulge. On the other hand, the simulated galaxies have 
spheroidal components that are roughly consistent with observations 
(\reffig{fig:mratio1}), suggesting that TNG100 generates similar but 
systematically less compact spheroids. It is unclear to what extent 
these results are affected by our neglecting the central 1 kpc region 
(because of resolution effects), and a fuller investigation must await 
analysis of higher resolution simulations. 

\subsubsection{Kinematic decomposition}
\label{sec:kimdecom}

\begin{figure*}[htbp]
\begin{center}
\includegraphics[width=0.95\textwidth]{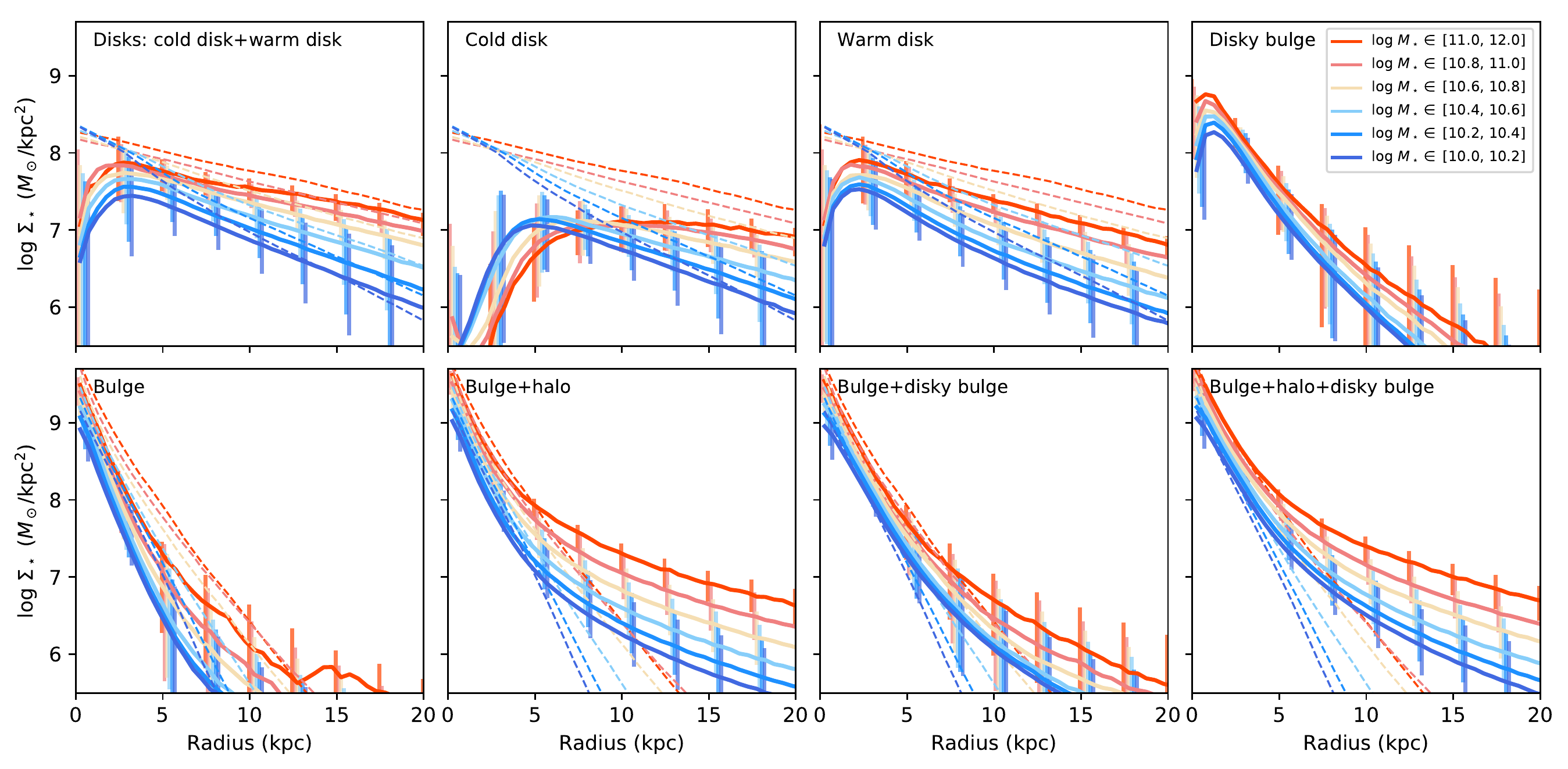}
\caption{
Radial surface density profiles of kinematic structures identified by \autoGMM, in $z=0$ disk galaxies from TNG100. We use the same sample as in \reffig{fig:SurfaceDensity2}, but adopt classification 2, and hence disky bulges are separated from disks. The solid profiles represent the median, and the error bars are the $1\sigma$ scatter. For comparison, we overlay the density profiles of the morphological disks (top panels) and bulges (bottom panels). In the bottom panels, we compare, from left to right, the surface density distribution between the morphological bulge with the kinematic bulge, the composite of bulge+halo, bulge+disky bulge, and bulge+halo+disky bulge. Clearly, the inner part of composite structure of bulge+halo+disky bulge best matches the morphological bulge.
}
\label{fig:SurfaceDensity3}
\end{center}
\end{figure*}

Figures \ref{fig:SurfaceDensity2} and \ref{fig:SurfaceDensity3} examine the face-on surface 
density profiles for the kinematic structures, dividing the galaxies into six bins of total 
stellar mass. The median profiles from the morphological bulge+disk decomposition are 
overlaid for comparison. In \reffig{fig:SurfaceDensity2}, the density profiles of disks 
(upper-left panel) are obtained by summing all kinematically disky structures (the cold and warm 
disks identified by classification 1), while those of spheroids (bottom-left panel) 
include both the halos and bulges. Both cold disks and warm disks (top panels) follow nearly exponential 
profiles with classification 1. However, cold disks are truncated in their inner regions, while 
warm disks are overmassive in their inner parts. This is consistent with the morphologies 
of the examples shown in \refsec{class1}. Some cold disks have similarly central 
overmassive components as well. As discussed in Section \ref{class2}, this is a clear 
imprint of the presence of disky bulges. 

As suggested in \refsec{class2}, it is necessary to separate disky bulges from 
other disky structures in order to explain the extra central concentrations. 
The location of disky bulges in the distribution of kinematic moments is 
marked by the dotted square in \reffig{fig:kinemphase}. Once we consider the 
disky bulge as a new, separate structure, the surface density profiles of cold 
and warm disks become simpler (\reffig{fig:SurfaceDensity3}). The warm disk 
follows a similar exponential profile as the cold disk, but it is clearly more 
centrally concentrated, with a smaller central 
truncation\footnote{\citet{Zhu2018b} also reported similar central truncations 
of disk orbits in the galaxies of the CALIFA survey.}. There are no central
overmassive components present for cold and warm disks in galaxies of different
stellar masses. The combined surface densities of cold and warm disks 
(upper-left panel of \reffig{fig:SurfaceDensity2}) also follow exponential 
profiles. From a kinematic point of view, disky bulges are essentially disky structures.
They rotate like warm disks. However, from the perspective of morphology, disky
bulges are more similar to bulges (upper-right panel of 
\reffig{fig:SurfaceDensity3}). Both of these traits are reminiscent of 
pseudo bulges \citep[e.g.,][]{Kormendy&Kennicutt2004}. 

\begin{figure*}[htbp]
\begin{center}
\includegraphics[width=0.95\textwidth]{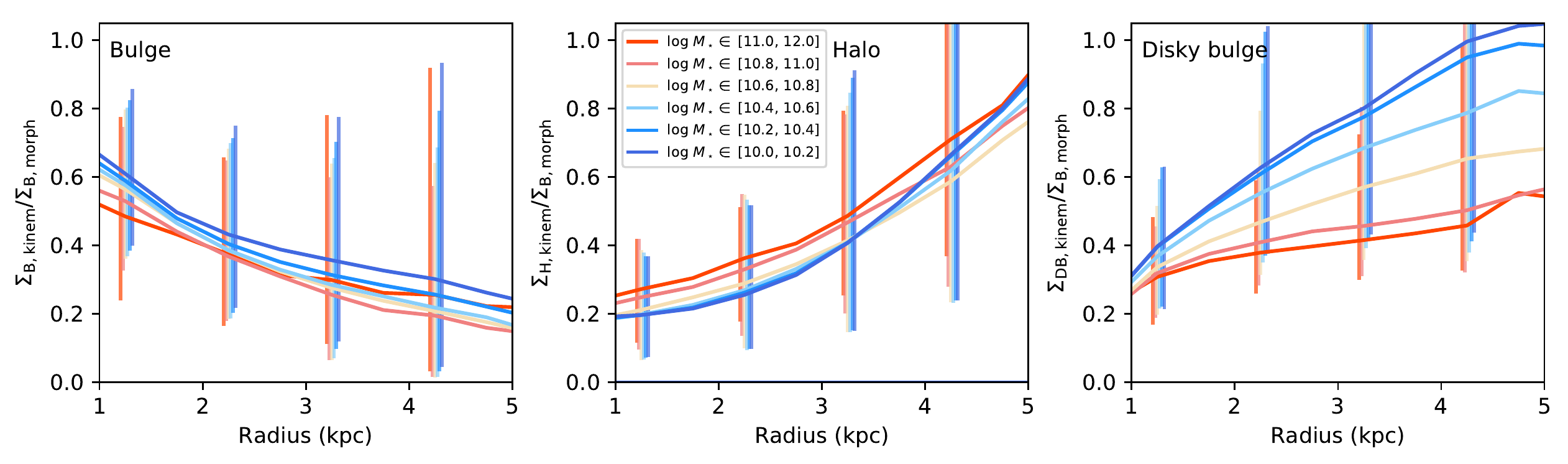}
\includegraphics[width=0.95\textwidth]{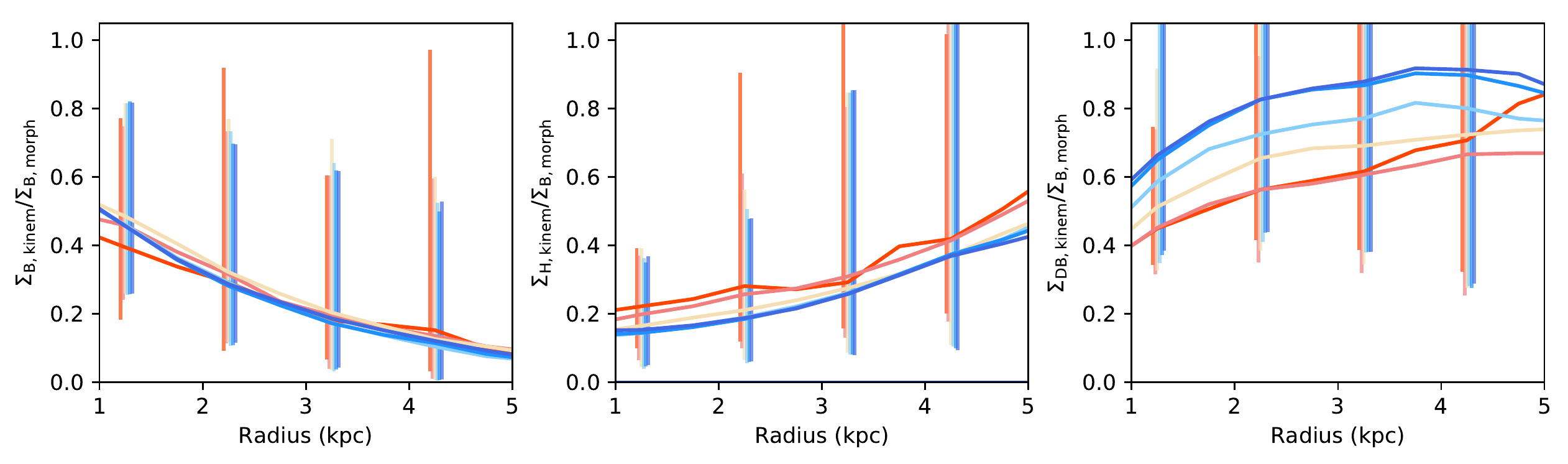}
\caption{Fractional radial deviation of morphological structures from kinematic structures identified by \autoGMM\ via classification 2 for different mass bins, for bulges (B), halos (H), and disky bulges (DB), from left to right. The denominators are the surface densities of morphological bulges (B, morph). The top and bottom panels are measured in the face-on and edge-on views, respectively. The solid profiles represent the median, and the error bars are their $1\sigma$ scatter. The values within 1 kpc are ignored.}
\label{fig:SurfaceDensity_4ratio}
\end{center}
\end{figure*}

The density profiles of bulges and halos are exactly the same in classifications 1 and 2, shown in 
the bottom panels of \reffig{fig:SurfaceDensity2}. The spheroidal structures of bulge+halo does not match 
morphological bulges well. In order to reconstruct morphological bulges, we show all 
possible combinations of the kinematic structures obtained by classification 2 in the 
bottom panels of \reffig{fig:SurfaceDensity3}. It is clear that morphological bulges match 
best the inner region ($R\lesssim 5$ kpc) of the case of bulge+halo+disky bulge. 
This suggests that morphological bulges are composite structures of bulges, 
halos, and disky bulges decomposed by kinematics. The relative contribution of these three 
components to morphological bulges is shown in
\reffig{fig:SurfaceDensity_4ratio}. Clearly, kinematic bulges dominate in the very central regions. 
The importance of disky bulges and halos increases outward. In face-on views (top panels), 
at $R\approx 2$ kpc the disky bulges have a comparable 
importance with kinematic bulges, then start to dominate toward larger radius. The 
kinematic halos contribute about one-third of the mass of morphological bulges at $R \leqslant 3$ kpc, 
and increases slowly outward. The same ratios but viewed edge-on are given in the bottom 
panels of \reffig{fig:SurfaceDensity_4ratio}. The disky bulges, that are flattened structures, 
become more prominent due to the projection. The contribution of kinematic bulges 
reduces to $\sim0.45$ at $R=1$ kpc. 

Our result is somewhat consistent with the well-known idea that bulges have two subtypes 
(classical and pseudo). But classical bulges seem to co-exist with both disky 
pseudo bulges and halos. Thus, the intrinsic kinematic structures may be significantly
mixed from a morphological point of view. Recently, some composite bulge systems have 
been found via detailed morphological decompositions of nearby galaxies \citep{Nowak2010, Mendez-Abreu2014, Erwin2015b}. 
Generally, compact, dynamically hot inner classical bulges were found to be embedded in 
pseudo bulges. The S\'ersic indices of such classical bulges are generally small ($n<2$) \citep{Erwin2015b}. 
These are qualitatively consistent with the composite bulges we obtain in \TNG\ galaxies. 

\section{Conclusions}
\label{conclusion}

\citet{Du2019} presented an automated method, \autoGMM, that can decompose galactic
structures efficiently for simulated galaxies based on in their kinematic phase-space properties. 
A few examples covering a broad range in mass and morphology show that \autoGMM\ 
recovers reasonable structures in an unsupervised way. Here, we apply this method to a large 
sample of present-day galaxies from the cosmological simulation \TNG. 
A sample of about 4000 disk galaxies of stellar mass 
$10^{10-11.6}\,M_\odot$, one-fourth of which are barred, is selected from TNG100
to analyse the statistical properties of structures identified by \autoGMM\ and to provide general 
insights about the relationship between kinematically and morphologically derived galactic structures. 
Such structures fall into a 
clear pattern in the kinematic phase space of normalized azimuthal angular momentum (circularity), 
non-azimuthal angular momentum, and binding energy. Thus, they are likely to be intrinsic structures. 

Two methods are introduced to classify the kinematic structures. The so-called ``classification 2'' 
gives five kinds of intrinsic kinematic structures: cold disks, warm disks, disky bulges, bulges, and 
halos whose kinematic and morphological properties are qualitatively consistent with thin disks, 
thick disks, pseudo bulges, classical bulges, and halos defined by morphology, i.e., profile fitting, 
in observations. 
We define cold and warm disks as disky structures having strong and moderate rotation, respectively. 
The stars of both bulges and halos distribute in spheroidal morphologies, but those of bulges are 
clearly more tightly bound. In ``classification 1'', disky bulges are considered 
as a part of disks that generally have a similar rotation to warm disks; however, their compact morphologies 
motivate us to separate them from disky structures in classification 2. 

Across all TNG100 $z=0$ disk galaxies of $10^{10-11.6}\, M_\odot$, the kinematic 
spheroidal structures, obtained by summing up stars of bulges and halos, contribute about 
$\sim 40\%-50\%$ of the total stellar mass within $3r_e$; the disky structures (cold disks, 
warm disks, and disky bulges) make up the remainder. The mass fraction of kinematic 
spheroids is consistent with the mass fraction of bulges decomposed via the morphological 
method. However, a systematic comparison between morphological and kinematic structures shows that they have significant differences in their morphologies. 

The bulges decomposed by morphology exhibit a composite structure that includes $\sim 60\%$ 
kinematic bulges, $\sim 20\%$ halos, and $\sim 30\%$ disky bulges at $R \approx 1$ kpc
in face-on view. At $R\approx 2$ kpc, disky bulges become equally important ($\sim 40\%$ of 
total stellar mass) with kinematic bulges in morphological bulges. 
Moreover, disky bulges are essentially disky structures that have a similar rotation to warm 
disks, and are likely to be classified as pseudo bulges in observations. Furthermore, our results
indicate that classical bulges commonly co-exist with disky pseudo bulges. 

The mixtures of multiple structures are less substantial in disks. Warm disks are generally more 
centrally concentrated than cold disks. The most surprising result is that most cold disks and 
many warm disks are sharply truncated in the galaxy central regions. The truncation radius of 
cold disks generally happens around $5$ kpc in lower mass galaxies, while it happens at larger radii
in massive ones. A similar phenomenon can be seen in a high resolution galaxy simulation of
the Milky Way \citep[][Figure 2]{Buck2019}. This suggests that the disk mass of the inner regions 
of galaxies is possibly 
significantly overestimated by a simple exponential fitting that is widely used in observations. 
Similarly, \citet{Zhu2018b} concluded that the central region of spiral galaxies are rather 
dominated by warm components than cold or hot components. Furthermore, we find that 
kinematic halos contribute $\sim 30\%$ of the mass of the outer part of morphological disks 
in face-on view. 

Galaxy structures identified via morphological methods (e.g. profile fitting of the photometry) may not reflect intrinsic galaxy components because of the underlying complexity of the stellar populations, the reduced information content of the images e.g. with respect to kinematics constraints, and the risks of human bias. Cosmological simulations of galaxies are a powerful tool to identify galaxies intrinsic structures and subcomponents and to relate them to physical processes. However, in order to be a reliable source of insights, the simulations need to return galaxies whose structures and properties are as realistic as possible. Despite the remarkable success achieved by TNG100, it remains challenging to make qualitative and quantitative comparisons with observations in detail, largely due to the resolution limitation. In future work, we will extend this analysis to high redshifts and the highest resolution run of the IllustrisTNG series \citep[TNG50][]{Nelson2019b, Pillepich2019}. The formation and evolution of each structure found in this work will be studied in greater detail.

\begin{acknowledgements}
This work was supported by the National Science Foundation of China (11721303, 11991052) and the National Key R\&D Program of China (2016YFA0400702).
M.D. is also supported by the National Postdoctoral Program for Innovative Talents (8201400810) and the Postdoctoral Science Foundation of China (8201400927) from the China Postdoctoral Science Foundation. V.P.D. was supported by STFC Consolidated grant ST/R000786/1. V.P.D. acknowledges support from the Kavli Visiting Scholars Program for a visit to the KIAA during this work. D.Y.Z. acknowledges the support by the Peking University Boya Fellowship. The TNG100 simulation used in this work, one of the flagship runs of the IllustrisTNG project, was run on the HazelHen Cray XC40-system at the High Performance Computing Center Stuttgart as part of project GCS-ILLU of the Gauss Centres for Supercomputing (GCS). The authors thanks all the members of the IllustrisTNG team for making the IllustrisTNG data available to us prior to their public release. We also thank Ling Zhu and Dandan Xu for constructive discussions. This work made use of the High-performance Computing Platform of Peking University. The analysis was performed using \texttt{Pynbody} \citep{pynbody}.
\end{acknowledgements}
 
\bibliographystyle{apj}
\bibliography{Reference_lib}

\end{document}